# Assessing Inconspicuous Smartphone Authentication for Blind People

**Diogo Marques**
University of Lisbon
dmarques@di.fc.ul.pt

**Luís Carriço**
University of Lisbon
lmc@di.fc.ul.pt

**Tiago Guerreiro**
University of Lisbon
tjvg@di.fc.ul.pt

**ABSTRACT**
As people store more personal data in their smartphones, the consequences of having it stolen or lost become an increasing concern. A typical counter-measure to avoid this risk is to set up a secret code that has to be entered to unlock the device after a period of inactivity. However, for blind users, PINs and passwords are inadequate, since entry 1) consumes a non-trivial amount of time, e.g. using screen readers, 2) is susceptible to observation, where nearby people can see or hear the secret code, and 3) might collide with social norms, e.g. disrupting personal interactions.

Tap-based authentication methods have been presented and allow unlocking to be performed in a short time and support naturally occurring inconspicuous behavior (e.g. concealing the device inside a jacket) by being usable with a single hand. This paper presents a study with blind users (N = 16) where an authentication method based on tap phrases is evaluated. Results showed the method to be usable and to support the desired inconspicuity.

**Author Keywords**
Tapping, inconspicuous interaction, smartphones, usable security, blind users.

**ACM Classification Keywords**
H.5.2 [**Information interfaces and presentation**]: User Interfaces – Input devices and strategies

**INTRODUCTION**
The growing capabilities and increasing adoption of smartphones is transferring privacy and security risks from our desks to our pockets. A typical way to mitigate these risks is to set up an authentication barrier that has to be overcome after a period of device inactivity. The unlocking procedure comes, however, with costs. Unlocking takes time, which must be multiplied by the number of occasions one pulls the device from the pocket. Additionally, authenticating creates an interruption that obstructs a primary objective, e.g. reading an email, thus potentially causing frustration. Finally, mobile authentication is particularly susceptible to observation attacks – a third party discerning one's secret code – since devices are often used outside controlled environments like home or office.

To blind users, authenticating with touch-based smartphones raises even more usability and security barriers. Let us consider using a PIN with a screen reader, such as iPhone's VoiceOver[1]. As the user passes its fingers through the screen, a voice reads out the key. When the right key is found, the user can split or double tap it to select. Azenkot et al. [2] found that even in a sample of blind users familiarized with this technology, unlocking took in excess of 7 seconds, which may be too cumbersome for large-scale adoption. Furthermore, it is susceptible to observation: the process of selecting each digit is visible to bystanders if a method to hide the screen is not used (e.g., iPhone's screen curtain). This can be particularly troublesome if we consider situations where blind people are less aware of their surroundings than sighted people. Aural eavesdropping is also a risk – it is conceivable that blind people could use headphones, but that would not only require time to put them on but also reduce awareness of surroundings. Finally, limitations imposed by social norms should also be considered – using visible authentication mechanisms can be embarrassing or raise fears of being perceived as distrustful of others.

In this paper, we contribute with a study on inconspicuous authentication, based on tap phrases, for blind people. Besides feasibility and performance, we report strategies mentioned by participants on how these methods could be used to mitigate security threats.

**RELATED WORK**
Tap-based authentication has been previously addressed in the literature. Bianchi et al. [2] propose using taps to insert PINs recurring to the number of touch events. PassChords [1] is specifically targeted to blind users. In this system, multi-finger touches, using a chord analogy, are used. These works do not allow unbounded tap pattern lengths, nor take into consideration the tap and break time spans, instead being focused on tap counts and location. The first severely reduces the tap phrases selection space. The second is prone to screen smudge detection and most notably limits the space for inconspicuous interaction strategies. For example, it does not afford operations with a single hand, rendering the most accepted concealing strategies impossible. Forcing authenticating in conspicuous ways draws attention to the use of assistive technology, potentially leading to feelings of self-consciousness [7].

---

[1]  http://www.apple.com/accessibility

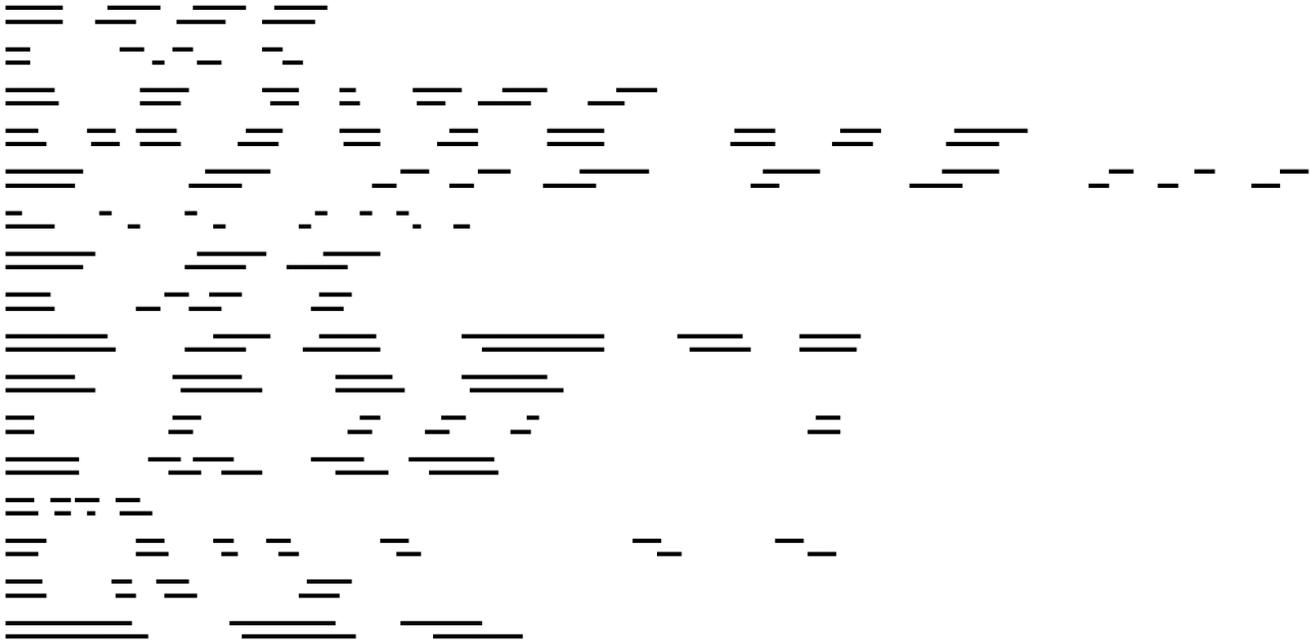

**Figure 1. Tap Phrase Authentication for all 16 participants. First line shows the recorded template and the next one shows the attempt to replicate it by the same participant.**

In TapSongs [8] users can configure tapping patterns representing songs. However, it uses a parametric approach, requiring at least 5 examples for configuration. For gestures, Li [4] points out that, only very few examples should be needed for subsequent accurate recognition. For taps, the case is even more evident. Using many examples and parametric approaches is prone to create a build-up of noise that prevents keeping the success rates and/or performance low, especially in low-end devices. Moreover, it discourages the configuration of long and securer codes. Also, to our knowledge, TapSongs has not been implemented in smartphones or similar devices, turning the user studies less realistic.

Ghomi et al. [3] attempt to generalize rhythmic authentication into a more general input method based on rhythm, making use of both taps and breaks. In this work the types of words that can exist are bounded to 5 – three varieties of tap and two of break, varying in canonical duration. Again the tap phrase selection space is limited. In our previous work [5], we devised a similar approach but enabling sequences of unbounded length and requiring only one template to record them.

Tap-based authentication has been previously presented as a method that enables non-visual usage. Still, particularly methods that allow concealing and one-hand usage, they have not been evaluated with blind people. It is paramount to understand how blind people cope with these tapping methods as well as their acceptance and concealing strategies.

**USER STUDY**
To validate tap phrase unlocking as a feasible approach to address the threats posed by traditional authentication, an exploratory user study with 16 blind participants was conducted. We addressed both the experience of using tap patterns for unlocking and the affordance of this method to inconspicuous interaction. Additionally, we extracted the patterns chosen by the users to develop an efficient matching algorithm, which we describe in section 5.

Specifically, our objectives were to understand if, after a short learning period, users could: 1) Perform authentication with a single hand easily and in a reasonable amount of time; and 2) confabulate strategies for inconspicuous authentication.

**Authentication method**
The tap-based authentication method used in this study was the one presented in [5]. *This method uses tapping sequences on a touch screen. The Hamming distance-based matching approach accommodates rich tapping patterns of unbounded length, taking into consideration both the times where the user is pressing and the ones in which she releases. Additionally, it tries to match every pattern candidate as the user interacts with the device, thus dispensing timeouts for successful authentication.*

**Materials**
For the experiments, a single Samsung Galaxy mini smartphone, with Android 2.3, was used. The prototype application for tap unlocking was used. A training mode was available in which data was not recorded and optional sound output (emission of a tone while the screen was being touched) was available. On data-recording mode, there were facilities for configuring a tap pattern template and afterwards to attempt unlocking, both without audio feedback. A short vibration was emitted on successful unlock. Success was determined by two factors: 1) the input and template having the same number of taps and 2) the input having total time span no more or less than 20% that of the template. This crude algorithm, not suited for authentication, was used so that users could have a more realistic understanding of the system. Logs were generated for offline analysis.

Paper questionnaires were employed to gather demographic data, register responses to the Single Ease Question (SEQ), and record concealment strategies suggested by participants. The SEQ, proposed by Sauro and Dumas [6], is known to have good psychometric properties, and was administered immediately after the task ("Overall, this task was?", Very Easy=7, Very difficult=1).

**Participants**

The 16 participants were volunteers recruited in a local training institution for blind people. Two participants had some residual vision. Ages varied between 26 and 64 years old, the average being 47 (SD = 12). Twelve participants were male and 4 female. Although all participants had mobile phones, they reported having none or very little experience with touch-screen devices. Eleven reported being very familiar with using PINs in electronic devices, albeit in physical keyboards. Participants reported never or rarely using headphones paired with their mobile devices.

**Procedure**

Participants were initially introduced to the concepts and explained the tasks they were asked to perform. At this stage, they were given no mention that the tap unlocking was method being proposed by the researchers. They were handed a device to feel and get accustomed to while being administered a short demographic questionnaire.

In a first stage, a training session lasting approximately 5 minutes was conducted, in the following steps:

1. Users freely explored the touch-screen area with their fingers. When they touched any point in the screen, an audio tone was emitted. Participants were explained that the whole screen acted as a single button and explored the fact that tap phrases are composed of taps and breaks lasting in time.

2. Users were asked to imagine tap phrases that they could record and later use for unlocking. They experimented freely, with sound enabled, until they were confident that they had grasped the concept.

3. Users were introduced to the vibrotactile feedback emitted on authentication success (short) and failure (long).

4. With sound output now removed, users conducted a complete dry-run, first configuring a template of their choice, then attempting to unlock.

5. Step 2 was repeated, this time participants using the device with a single hand.

After training, still using a single hand, participants were asked to again configure a template and then try to unlock. This time, the interaction was recorded in log files. Immediately after completing this task, users responded to the Single Ease Question.

In the second stage of this study, participants were reminded of the observation threat and asked to imagine strategies they could use to conceal the input from potential observers. To facilitate this process, participants engaged in role-playing two scenarios: a meeting and a commute in public transportation. To that end, they were given a smartphone so they could simulate authentication. A facilitator gravitated at times around the participant to make him aware of possible visual observation angles. In the end, participants were asked to summarize the viable strategies they had identified.

**Measures**

For the unlocking task, we acquired:

1. The time it took to complete authentication with a single hand, measured from the moment a facilitator clicked a start button and initiated the unlocking application to the moment an input was accepted as the correct secret code;

2. The number of input errors, and;

3. The SEQ rating, from 1 to 7.

For the elicitation part of this study, the strategies indicated by participants were recorded and occurrences counted. Since the alternatives mentioned were clear and not very numerous, no special categorization was performed.

**Results**

After training, all users were able to authenticate in the first trial, so there were no input errors to record.

Figure 1 presents the recorded tap phrases along with the attempt to replicate it for all 16 users.

The mean task completion time was 4.32s, with standard deviation 2.1s. A Shapiro-Wilk test suggests that the data is normally distributed (S-W = .890, df = 16, p = .056). A one-sample t test was conducted at an alpha level of .05 to evaluate if tap unlocking is faster than the 7.52s mean for PIN with VoiceOver found by Azenkot et al. [1]. The test has shown to be significant (t = -6.062, df = 15, p = .000, Cohen's d = -1.52), suggesting that unlocking with taps is faster than the current mainstream method (Voice Over).

The median SEQ score was 6 (IQR = 3), where 7 means "very easy" and 1 "very difficult", indicating that participants perceived tap unlocking as being easy to perform.

In the second stage of our study, we elicited inconspicuous authentication strategies through role-playing. The user-suggested approaches are summarized in Table 1. Each user contributed, on average, 3 strategies (SD = 1). The top suggestions, with 9 occurrences, were performing authentication under the table or inside a pocket.

**Discussion**

The results for task completion time and perceived easiness of authenticating with tap phrases are encouraging. Even so, the relatively large standard deviation in task completion time deserves a closer look. From our observations, there are two possible explanations for this fact: 1) some users, lacking the confidence and experience using smartphones, operated the device with an unusual level of caution, thus taking more time and 2) there may be, in fact, an extended initial period where a blind user needs to situate himself before starting tapping with confidence.

| Strategy | Occurrences |
|---|---|
| Under the table | 9 |
| Inside pocket | 9 |
| Occluded by clothes (e.g. jacket) | 7 |
| Cover with one hand | 5 |
| Lean device against body | 3 |
| Inside bag / purse | 3 |
| Using device upside down | 3 |
| Move to an isolated location | 2 |
| Under the seat | 1 |
| Postpone | 1 |

**Table 1.** Concealing strategies suggested by study participants.

The top suggestions for inconspicuous authentication strategies include many cases that are made possible, or at least easier, by the tap phrase method. This is true not only for blind users, but in any situations where the visual channel is not available. Our emphasis focus on affording single hand interaction required by some of top suggestions, e.g. authenticating inside the pocket or using the free hand to cover the interaction. The feasibility of actually using some of the selected strategies must, however, be further evaluated in realistic settings. For example, using a pocket may not be possible because hand movements can be constrained. That's not the case for most of the others.

**CONCLUSIONS**

Mobile devices are increasingly seen as extensions of ourselves, permeating our personal and social life. But they also create new threats to our well-being that can have devastating effects. The ability to self-protect is a universal requirement in human endeavors, from which the ubiquitous computing enterprise cannot escape if it is to be successful and inclusive. In this paper, we presented a novel approach for mobile authentication that attempts to foster inclusion of blind people in the global trend towards smartphone adoption.

We propose a method in which authentication is accomplished by recognizing rich tapping patterns that can be performed with a single hand, using a smartphone's entire screen as a single button. We presented results of a user study that shows that this method is usable and that it affords inconspicuous interactions, thus not only offering increased security but also enabling compliance with social norms.

Clear avenues for further research were opened, namely exploring longer-term usage and real-world feasibility of inconspicuous authentication scenarios, measuring resistance to eavesdropping, and expanding the matching of tap patterns to accommodate other interactions other than authentication.

**ACKNOWLEDGMENTS**
We thank Fundação Raquel and Martin Sain and the participants of the study.